\documentclass[aps, reprint, superscriptaddress, nofootinbib]{revtex4-1}
\pdfoutput=1
\usepackage{amsmath, latexsym, amssymb, hyperref, graphicx, color, slashed}
\definecolor{nicered}{rgb}{0.7,0.1,0.1}
\definecolor{nicegreen}{rgb}{0.1,0.5,0.1}
\hypersetup{colorlinks, citecolor=nicegreen,linkcolor=nicered}
\begin{document}

\title{
   Restoration of Parity and the Right-Handed Analog of the CKM Matrix
 }

\author{Goran Senjanovi\'{c}}
\affiliation{Gran Sasso Science Institute, Viale Crispi 7, L'Aquila, Italy}
\affiliation{International Centre for Theoretical Physics, Trieste, Italy }
\author{Vladimir Tello}
\affiliation{Gran Sasso Science Institute, Viale Crispi 7, L'Aquila, Italy}
\date{\today}

\begin{abstract}
 
    In a recent Letter we determined analytically the right-handed quark mixing matrix in the minimal Left-Right symmetric theory with generalized Parity. We derived its explicit form as a series expansion in a small parameter that measures the departure from  hermiticity of quark mass matrices. Here we analyze carefully the convergence of the series by 
    including 
     higher order terms and by comparing with   numerical results. 
    We apply our findings to some
    phenomenological applications such as the production and decays of the right-handed  gauge boson $W_R$, the neutrinoless double beta decay,  the decays of the heavy scalar doublet, the strong CP parameter and the theoretical limits on the new mass scale from the $K$ and $B$-meson physics. In particular, we demonstrate that the relevant coupling for the production of the $W_R$ gauge boson at  hadronic colliders and for the neutrinoless double beta decay  
    equals its left-handed counterpart, within a percent. 
    We also demonstrate that the stability of the theoretical lower limit on the $W_R$ mass
  from the $K$-meson physics
     is due to a partial cancellation of the external phases of the right-handed mixing matrix.
    
    \end{abstract}

\maketitle

\section{Introduction} 

    The Left-Right symmetric theory~\cite{lrmodel}  was put forward forty years ago as an attempt to understand the origin of parity violation in weak interactions through the spontaneous symmetry breaking. The left-right (LR) symmetry was chosen originally as a generalized parity ($\mathcal{P}$), although it could also be a generalized charge conjugation  ($\mathcal{C}$). Regardless of the nature of LR symmetry, the theory leads to a non-vanishing neutrino mass and to the seesaw mechanism~\cite{Minkowski:1977sc,Mohapatra:1979ia,seesawso10,seesawfam}.
     The smallness of neutrino mass is related to the departure from the maximal  violation of parity at low energies~\cite{Minkowski:1977sc,Mohapatra:1979ia}. 
 Furthermore, the knowledge of neutrino masses and mixings allows to predict their Dirac Yukawa couplings and the associated physical phenomena~\cite{Nemevsek:2012iq}.

   Once the left-right symmetry is broken, the question is raised regarding the connection between the right-handed (RH) and left-handed (LH) quark and lepton mixing matrices. In the leptonic sector this connection goes away due to the seesaw mechanism which
   differentiate left and right handed neutrinos, while in the quark sector the situation depends on the nature of LR symmetry. 
   
   In the case of $\mathcal{C}$, symmetric Yukawa couplings lead to symmetric quark mass matrices and thus to  same left and right mixing angles,  same magnitude but opposite signs of the Kobayashi-Maskawa (KM) phase and its right-handed analog, and to arbitrary external phases in the RH sector. 
   The equality of mixing angles imply the same strength for production and hadronic decay rates of left and right-handed charged gauge bosons, while the new RH phases bring in new sources of CP violation. 
   
   In the case of $\mathcal{P}$, the situation is not that simple. In spite of having hermitian Yukawas couplings,  the spontaneous symmetry breaking of $\mathcal{P}$ introduces a complex phase which spoils the hermiticity of quark mass matrices, and a priori one would imagine no clear relation between LH and RH quark mixings. This soon became a burning issue, and the determination of the RH quark mixing matrix became over the years a kind of Holy Grail of the theory. Its explicit analytic form was only recently achieved~\cite{history} in the entire parameter space. In this longer sequel to our Letter we provide the missing technical details and the higher order terms in the series expansion of $V_R$. We also discuss some important phenomenological applications and their consequences.

   A few words regarding the history of the attempts to determine the RH quark mixings. The first step in this direction was made in Ref.~\cite{Kiers:2002cz} through a numerical study in a portion of the parameter space. Some years later, Ref.~\cite{md} made an  analytical study in the same approximation and found the same hierarchical structure as in the CKM matrix. Finally,   this result was established over the entire parameter space by combining analytical and numerical computations in Ref.~\cite{Maiezza:2010ic}. 
   What was missing until Ref.~\cite{history} was the exact equation for $V_R$, its analytical solution and a clear demonstration of the approximate equality of mixing angles of the left and right-handed sectors.  
   
     A few years ago, a general low energy study of the RH quark mixing matrix was performed in Ref.~\cite{Blanke:2011ry}. 
     More recently, it was argued that the near equality of the RH and LH mixing angles can be studied at the LHC with the b-quark tagging, and probed with a good precision at high luminosity~\cite{Fowlie:2014mza}, providing yet another motivation for our in-depth study of the RH mixing. 
 
 The knowledge of the RH quark mixing is crucial for a number of phenomenological reasons, both at low energies and at the hadronic colliders.

 First and foremost, the Majorana nature of left handed and heavy right handed neutrinos leads to neutrinoless double beta decay \cite{Racah:1937qq} through both left and right-handed gauge interactions~\cite{Mohapatra:1979ia,Mohapatra:1980yp}. The RH contribution could easily dominate over the usual LH one, providing a natural example of new physics being responsible for this process~\cite{Feinberg:1959}.
 
 Moreover, one can in principle observe lepton number violation at hadronic colliders in the form of same sign charged di-leptons, and probe directly the Majorana nature of heavy neutrinos through the so-called Keung-Senjanovi\'{c}  (KS) process~\cite{Keung:1983uu}. Its potential discovery would allow for precise predictions~\cite{Tello:2010am} (see also~\cite{Chakrabortty:2012mh}) for neutrinoless double beta decay and lepton flavor violation; 
 see e.g.~\cite{Senjanovic:2010nq} for a review. Detailed studies~\cite{Ferrari:2000sp} support the feasibility of the KS process at the LHC and a possible probe of the LR scale all the way up to 5-6 TeV; for a roadmap, see~\cite{Nemevsek:2011hz}. Recently the LHC has set the lower limit on the LR scale limit of roughly 3 TeV for a wide rage of RH neutrino masses~\cite{Khachatryan:2014dka}.
 It should be stressed that it is possible to determine the chirality of $W_R$ gauge boson couplings, as discussed in the KS~\cite{Han:2012vk} and the hadronic channels~\cite{Gopalakrishna:2010xm}. 
 
        While the hadronic colliders may be the best for a discovery of $W_R$ due to their high energies, it is important to study also other collider possibilities. For a linear collider analysis we refer the reader to~\cite{Huitu:1999qx}.
        Recently, a study of the theory for the $e p $ colliders was done in~\cite{Kaya:2015tia}.

    It was known for a long time that the small $K_L - K_S$ mass difference leads to the lower limit on the LR scale in the minimal model~\cite{soni}, around 3 TeV (for the most recent study, see~\cite{Bertolini:2014sua}). In the case of $\mathcal{P}$, from the study of the electric dipole moment of the neutron, the limit was raised to about 7 TeV~\cite{Xu:2009nt},  without taking into account the strong CP violation. Meanwhile new studies also improved the chiral perturbation results for this process~\cite{Engel:2013lsa,Maiezza:2014ala} and recently~\cite{Maiezza:2014ala} claimed a limit (which depends though on the ultraviolet completion of the theory) of about 20 TeV by studying carefully the strong CP.

           It may be worth mentioning that the CMS reported recently~\cite{Khachatryan:2014dka} a 2.8 sigma excess in the KS process that can be interpreted as an indication of the LR symmetry~\cite{Deppisch:2014qpa}. It would require however the right-handed gauge coupling appreciably smaller that its left-handed counterpart, which would imply the breaking of LR symmetry at a very high scale. In what follows we work in the minimal model with single scale of symmetry breaking.

    The rest of the paper is organized as follows. In section~\ref{VR} we discuss our main results, i.e., the derivation of the RH mixing matrix as a perturbation series in a small parameter which measure the hermiticity of the quark mass matrices. 
    We show that mixing angles are well approximated already at the first order; the differences between RH and LH angles is in excellent agreement with the numerical results. The same is true for the KM phase and its RH counterpart, while the external phases in the RH mixing matrix need higher order corrections.  We then discuss the issue of the convergence of the external phases.

             The phenomenological applications of our results are left for  section~\ref{pheno}, where we discuss the production and decays of the RH charged gauge boson, the neutrinoless double beta decay, the decays of the heavy scalar doublet of the theory, the limits in the LR scale from the K and B meson physics and the computation of the strong CP parameter $\bar \theta$. 
               
               Finally, in section~\ref{s&o} we summarize our findings and offer an outlook for future work.
               
              In  Appendix~\ref{LO} we give the technical details necessary for the derivation of 
               the analytic expression of the RH mixing matrix at the first order in perturbation theory. We discuss there how to deal with the square roots of matrices needed for this. In  Appendix~\ref{HigherO}  we derive the  the second and third order terms of the same expansion.
                 In  Appendix~\ref{0mixing}, for illustrative purposes, we give exact values for the external phases in the case of vanishing mixing angles.                            

Before we turn to the task of computing the RH mixing matrix, a comment is called  regarding the question of potentially dangerous domain walls~\cite{Zeldovich:1974uw}, the product of a spontaneous breaking of discrete symmetries. A natural way of avoiding them is of course inflation, but in a low scale  theory it is not available. Another natural way~\cite{gia} may be non-restoration of symmetries at high temperature~\cite{nonrest}, but even that may not work~\cite{Bimonte:1995xs}. In any case, a tiny (even Planck scale suppressed) explicit breaking~\cite{balram} suffices to get rid of domain walls.

\section{The right handed quark mixing matrix}
\label{VR}
   The LR symmetric theory studied here is based on the $SU(2)_L \times SU(2)_R \times U(1)_{B-L}$ gauge group augmented with  generalized parity $\mathcal{P}$: $q_L \leftrightarrow q_R$ (for recent reviews, see ~\cite{Senjanovic:2010nq}).
Quarks and leptons come in LR symmetric representations
\begin{equation}
  Q_{L,R } = \left( \begin{array}{c} u \\ d \end{array}\right)_{L,R},\qquad
  \ell_{L,R} = \left( \begin{array}{c} \nu \\ e \end{array}\right)_{L,R}.
  \label{ds21}
\end{equation}

  The Higgs sector consists of the following multiplets \cite{Mohapatra:1979ia}: the bi-doublet
$\Phi$ and the $SU(2)_{L,R}$ triplets $\Delta_L $ and $\Delta_R$,
where under generalized parity $\mathcal{P}$:~$\Phi \leftrightarrow \Phi^{\dagger}$ and $\Delta_L \leftrightarrow \Delta_R$.

At the first stage of symmetry breaking, the neutral component of $\Delta_R$ develops a vev and breaks the original symmetry down to the SM one. The latter is in turn broken through the vevs of the neutral components of $\Phi$ 
\begin{equation}
\langle\Phi\rangle=v\, \text{diag} (\cos\beta,-\sin\beta e^{-ia})
\end{equation}
where $v$ is real and positive and $ \beta<\pi/4$, $0 < a < 2 \pi$. 

   The quark Yukawa couplings in the minimal theory take the following form 
\begin{equation}\label{eq:quarks&phi}
- L_Y=  \overline{q_{L}}\,\big(Y_{1} \Phi - Y_{2}\,  \sigma_2 \Phi^* \sigma_2)\, q_R+\text{h.c.}
\end{equation}
which gives  the following mass matrices for quarks
\begin{equation}\label{eq:yuk1}
 \left(\begin{array}{c}
M_u \\[3pt] 
M_d
\end{array}\right) = v
\left(\begin{array}{c c}
- c_{\beta} &-e^{ia}s_{\beta}  \\ 
e^{-ia}s_{\beta} & c_{\beta}
\end{array}\right)
\left(\begin{array}{c}
Y_1 \\[3pt]
Y_2
\end{array}\right)  
 \end{equation}
where $c_{\beta}\equiv \cos\beta$ and $s_{\beta}\equiv \sin\beta$. The underlying
 generalized parity $\mathcal{P}$ implies hermitian Yukawa couplings
\begin{equation}\label{hermyuk}
Y_{1,2}^{\dagger} = Y_{1,2}
\end{equation}
 which in turn leads to the following relations between the up and down quark mass matrices
\begin{align}\label{relationsMuMd-1}
M_u-M_u^{\dagger}&=-i s_at_{2\beta} (e^{-ia}t_{\beta}M_u+M_d)\\ \label{relationsMuMd-2}
M_d-M_d^{\dagger}&=is_at_{2\beta}(M_u+e^{i a}t_{\beta}M_d)
\end{align}
where $s_a \equiv \sin a$, $t_\beta \equiv \tan \beta$, $t_{2\beta} \equiv \tan 2\beta$.
The amount of the hermiticity of quark mass matrices is measured by the $s_at_{2\beta}$ parameter.  From \eqref{relationsMuMd-2}, by focusing on the third generation and ignoring its mixings with the first two, it is straightforward to obtain a rough upper limit $s_a t_{2\beta} \lesssim 2 m_b/m_t$. We justify it through an exact result found below. 

From the hierarchy of top and bottom quark masses, it is easy to see that $M_u$ is almost hermitian.   Assuming  small $\tan\beta$,  Ref.~\cite{md} 
  worked
in the approximation of  hermitian $M_u$ and obtained a semi-analytical form of the RH quark mixing matrix. In what follows we derive an exact equation for $V_R$ in the full parameter space and find an analytic solution as a series in the small parameter $s_at_{2\beta}$.
 
To set notation
 \begin{equation}\label{notationUD}
  M_u=U_L m_u U_R^{\dagger},\quad M_d=D_L m_d D_R^{\dagger}
\end{equation}
where $m_q$ are diagonal matrices of positive quark masses. The left-handed CKM matrix $V_L$ and its right-handed analog $V_R$ are then given by
\begin{equation}
  V_L=U_L^{\dagger}D_L,\,\,\,  V_R=U_R^{\dagger}D_R
 \end{equation}
 as manifested in LH and RH charged gauge interactions 
\begin{equation} \label{eqWLWR}
  \mathcal L_{gauge} = - \frac{g}{\sqrt 2} \left( 
  \overline u_L V_{L}  \slashed{W}_{\!L} d_L + 
  \overline u_R  V_{R}  \slashed{W}_{\!R} d_R\right) + \text{h.c.}
\end{equation}

  We refer the reader to  Appendix~\ref{LO} for the  details that go into the following leading order expression~\cite{history}
\begin{widetext}
\begin{equation}\label{eq:master}
(V_R)_{ij}= (V_L)_{ij} - i   s_a t_{2\beta}\bigg[  t_{\beta}  (V_L) _{ij}
+ \frac{  (V_Lm_d V_L^{\dagger} )_{ik}  (V_L )_{kj} }{m_{u_i}+m_{u_k}}  +  \frac{(V_L)_{ik} ( V_L^{\dagger}m_uV_L)_{kj} }{m_{d_k}+m_{d_j}} \bigg]  
+O(s_a^2t^2_{2\beta}) 
\end{equation}
\end{widetext}
This is not the only solution; the others are found through $V_L\rightarrow S_uV_L S_d$ and $m_{q_i}\rightarrow s_{q_i} m_{q_i}$, where $S_u=\text{diag}(s_{u_i})$, $S_d=\text{diag}(s_{d_i})$ and $s_{q_i}$ are $\pm$ signs. This is discussed at the end of the Appendix~\ref{LO}. Notice that 
 even when quark mass matrices are hermitian, the LH and RH mixing matrices are not automatically equal due to the sign freedom of quark masses; instead one has the well known result $V_R=S_u V_L S_d$. 

Equation \eqref{eq:master}  gives
 the RH mixing matrix $V_R$ as an expansion in the small parameter $s_a t_{2\beta}$, which, for the sake of convergence, must satisfy 
\begin{equation}\label{eq:limit}
|s_{a}t_{2\beta}|\lesssim \text{min}  \bigg|\dfrac{ m_{d_i}+m_{d_j}}{(V_L^{\dagger}m_{u}V_L)_{ij}} \bigg| \simeq  2 \frac{m_b}{m_t}
\end{equation}
This agrees with what we estimated taking the third generation and should not come as a surprise since the third generation CKM mixing angles are basically negligible in this context. 

When $s_a t_{2\beta}$ is close to $ 2 m_b/m_t$ one should include higher order terms, which can be found in the Appendix~\ref{HigherO}.
 This turns out necessary for the phases, while the values of the RH mixing angles  obtained from \eqref{eq:master} agree very well with the numerical solution of the exact equation \eqref{eq:VR}.

The first
 term in the bracket in \eqref{eq:master} is quite small,
  while the third term generally dominates over the second due to large mass ratios of the top quark mass over the down quark masses (basically due to large $m_t/m_b$), and to some degree, the ratio of charm quark mass over $m_d$ and $m_s$.
 It is noteworthy that the above results are parametrization independent.
 
It is evident from \eqref{eq:master} that a diagonal $V_L$ implies a diagonal $V_R$ at this order. We will see that this fact persists at higher orders and, moreover, it is clearly a solution of the exact equation  \eqref{eq:VR}. The difference between RH and LH mixing angles is controlled by the smallness of the CKM angles and the smallness of the parameter $s_a t_{2\beta}$, as we discuss amply in what follows.

A comment is in order.  Our results, such as formula \eqref{eq:master}, are obtained by using generalized parity $\mathcal{P}$ and  are valid at their face value only at the  scale of LR symmetry breaking or above.  They can be safely applied at high energy; however, in order to use them at low energies one needs to take into account their scale dependence.  We are interested principally in a LR scale accessible to the LHC which makes these effects tiny and we will ignore them in the rest of this work.

\subsection{ RH angles and KM phase.}

We compute next the difference between the left and right mixing angles and the KM phase and its RH counterpart. To do this we use the  following parametrization for a general $ 3 \times 3$ unitary matrix  
\begin{equation} \label{pdg-parametrization}   
\!\!V_R\! \equiv\! \text{diag}(e^{i\omega_1}\!,e^{i\omega_2}\!,e^{i\omega_3}\!)\,V(\theta_{ij}^R,\delta_R)\, \text{diag}(e^{i\omega_4}\!,e^{i\omega_5}\!,1)
\end{equation}
where  $V(\theta^{ij}_L,\delta_L) \equiv V_L$   is the standard form of the left-handed CKM matrix used by the Particle Data Group. Besides the RH analog $\delta_R$ of the KM phase $\delta_L$ (from now on both called KM phases), $V_R$ contains five external phases which cannot be rotated away since we used all the phase freedom in defining the usual CKM matrix in the left sector. 

A straightforward computation from \eqref{eq:master} gives the leading terms for the differences between mixing angles
    \begin{align}\label{eq:difference12}
   \theta^{12}_R - \theta^{12}_L &\simeq  - s_a t_{2\beta} \frac {m_t}{m_s} s_{23} s_{13}   s_\delta
\\[3pt]
   \label{eq:difference23}
   \begin{split}
\theta^{23}_R - \theta^{23}_L &\simeq -  s_a t_{2\beta}  \frac{m_t}{m_b}\frac{m_s}{m_b} s_{12} s_{13} s_\delta
\end{split}
\\[3pt]
    \label{eq:difference13}
    \begin{split}
     \theta^{13}_R - \theta^{13}_L &\simeq - s_a t_{2\beta} \frac{m_t}{m_b} \frac{m_s}{m_b} s_{12} s_{23} s_\delta\end{split}
 \end{align}
 and similarly for the KM phases
\begin{align} 
   \label{eq:Delta}
  \delta_R - \delta_L  &\simeq  s_a t_{2\beta} \frac{m_cc_{23}^2 +m_t s_{23}^2}{m_s}
  \end{align}
 where, for simplicity, we defined $s_{ij}=\sin \theta^L_{ij}$, $c_{ij}=\cos \theta^L_{ij}$ and $s_\delta= \sin \delta_L$.
  These are  leading order terms, more complete expressions for the angles were given in~\cite{history}, and the exact first order terms can readily be obtained from \eqref{eq:master}.

 It suffices to change the signs of quark masses accordingly to get all the other solutions.  As shown below, the absolute values of the mixing angle differences are quite stable under these transformations, while the KM phase difference varies somewhat.
    
  Notice that the angle differences vanish in the limit of CKM phase $\delta_L$ going to zero.
This follows from the fact that  in this limit, the first order terms in $s_a t_{2\beta}$ in \eqref{eq:master} are purely imaginary and thus affect only the phases. From the above formulas, it is evident that the angle differences are very small, suppressed by small CKM mixings. Moreover, in the case of 2-3 and especially 1-3 mixing angles there is an additional suppression of a small quark mass ratio $m_s/m_b$ that compensates for the large $m_t/m_b$ factor. As if there was a conspiracy in nature to keep the symmetry between LH and RH mixing angles in a world with broken parity.
  
\begin{figure} 
  \hspace{-0.3cm}  \includegraphics[scale=0.515]{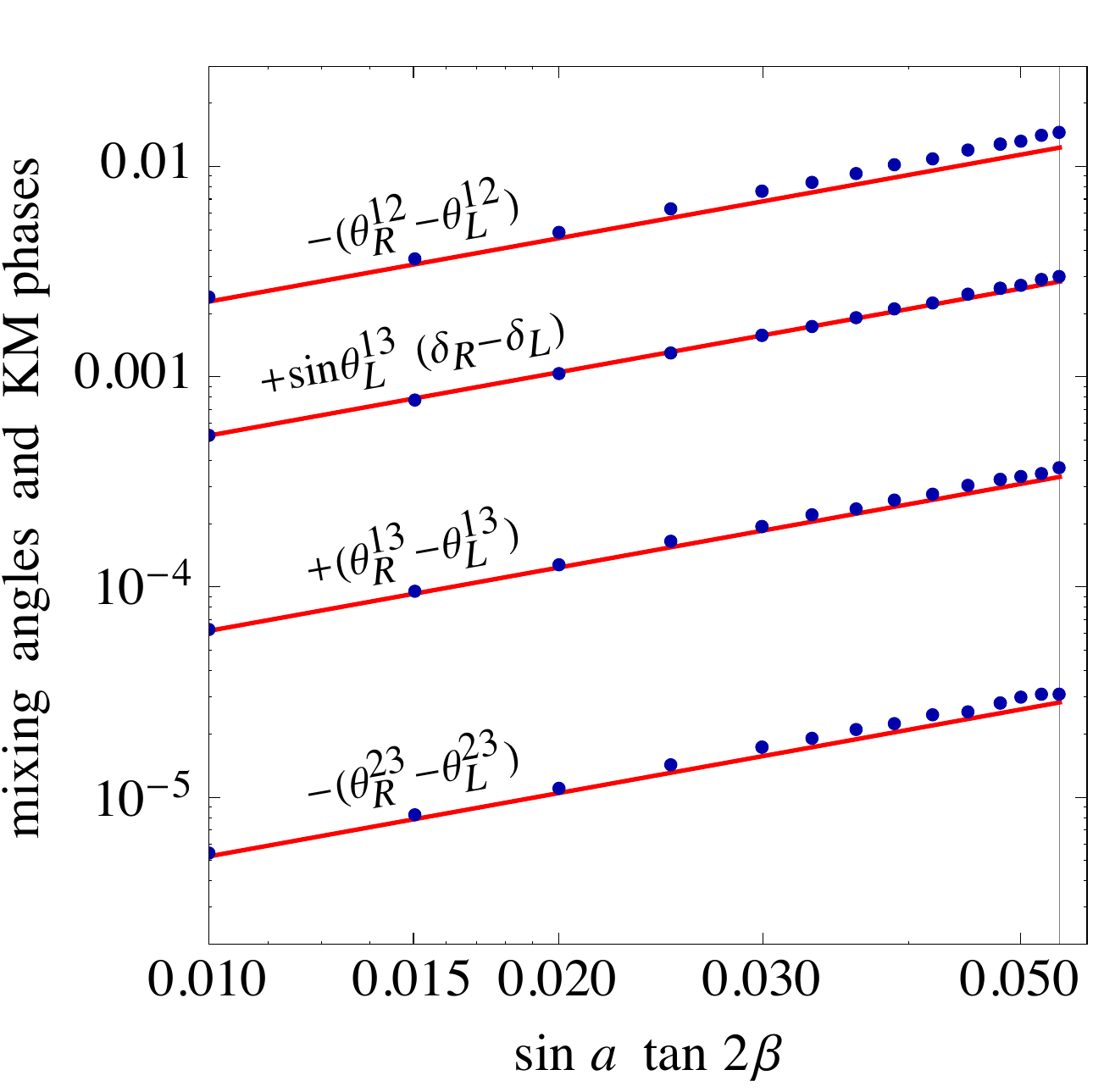} 
\caption{The differences between the right and left handed mixing angles and the KM phases. The first order terms are given by red lines, the blue dots denote numerical solutions of the exact equation. The agreement is manifest in the entire physical region $s_a t_{2\beta} \lesssim 0.54$. }
\label{fig:VR-angles}
\end{figure}

  In Fig.~\ref{fig:VR-angles}  we plot in red lines these first order results, and with blue dots the numerical solutions of the exact equation.   The first order is an excellent approximation and the agreement between the two is manifest through the  whole physical range of $s_at_{2\beta}$. Notice that the phase difference $\delta_R-\delta_L$ is multiplied  with the factor $\sin\theta^{13}_L$.  The reason is  that the   phases $\delta_L$ and $\delta_R$ are always accompanied with  $\sin\theta^{13}_L$ and $\sin\theta^{13}_R$ (which are practically the same), respectively. 
  
    The crucial point, as we noticed, is that the differences of  LH and RH mixing angles  are always proportional to another small LH mixings, which  control their smallness. From Fig.~\ref{fig:VR-angles} it seems that this holds true in higher orders of perturbation in $s_a t_{2\beta}$ and the proof comes from a discussion of a non-realistic two generation situation. Indeed, from \eqref{eq:master}, the difference of left and right mixing angles is zero, since $V_L$ is real. 
    In~\cite{history} we showed that the remarkable equality of $\theta_L$ and $\theta_R$ is actually exact in the two-generation case, which then guarantees the smallness of the mixing angles differences at all orders. 
 
\subsection{ RH external phases.}
 
First, we compute the external phases from \eqref{eq:master} and \eqref{pdg-parametrization}
 \begin{align}\label{eq:extphasesw1} 
\omega_1 \!&  \! \simeq   \! - \omega_3 \!+\! s_at_{2\beta}\bigg(\! \frac{m_c c_{23}^2 \!+\!m_ts^2_{23}}{m_s} 
   \! -\!\frac{m_dc_{12}^2\!+\!m_s s^2_{12}}{2m_u}\! \bigg) 
   \\[5pt] 
    \label{eq:extphasesw2} 
  \omega_2\!& \simeq - \omega_3 \simeq s_at_{2\beta} \frac{m_t}{2m_b}
        \\[5pt]
     \label{eq:extphasesw4} 
    \omega_4 \! & \simeq  \! \omega_3 \!- \!s_at_{2\beta}\bigg(\! c_{12}^2\frac{m_c c_{23}^2\!+\!m_ts^2_{23}}{m_s}\!+\!s^2_{12}\frac{m_c c_{23}^2 \!+\!m_ts_{23}^2}{2m_d}\!\bigg) 
   & \\[0pt]
    \label{eq:extphasesw5} 
     \omega_5\!& \simeq  \omega_3 - s_at_{2\beta} \, c_{12}^2\frac{m_c c_{23}^2 +m_ts^2_{23}}{2m_s}  
  \end{align}

The expressions above are somewhat more precise that what we gave in the previous Letter version of this work.
Unlike the expressions for the mixing angles and the KM phases, the external phases depend strongly on the sign transformations that connect different solutions. The above formulas should be taken as an example with all positive signs. It is straightforward to get more precise and complete expressions for all the cases. There is one subtlety to keep in mind: in some cases sign changes make the phases start from $\pi$ and not from zero, but that is easy to figure out.

\begin{figure}
\includegraphics[scale=0.5]{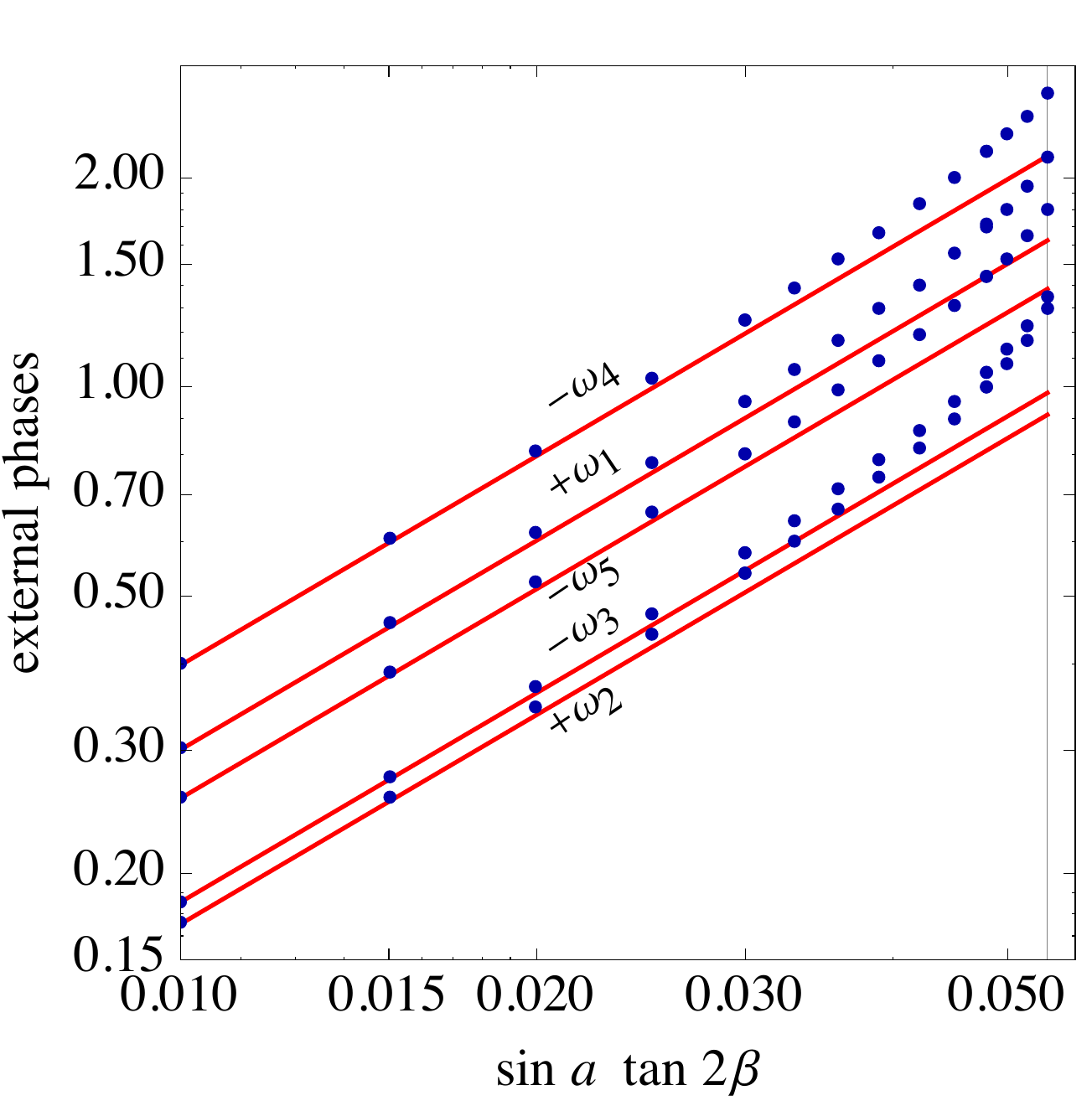}
\caption{Right handed external phases. In red lines we give the first order expressions and with blue dots the numerical solutions of the exact equation; they  start diverging only for larger values $s_a t_{2\beta} \gtrsim 0.03 $.}
\label{fig:VR-phases}
\end{figure}

 We plot these phases in Fig.~\ref{fig:VR-phases}. Again, the first order results are shown in red, and the numerical results in blue. Notice that in this case the results start diverging for larger values  $s_a t_{2\beta} \gtrsim 0.03 $, which simply implies the need for higher order terms in \eqref{eq:master}, as discussed below.
%

\subsection{The impact of sign alternatives}

The reader must be worried about the different signs that plague \eqref{eq:master}, due to sign freedom of quark masses. It is easy to see in the first order terms  that only the signs of mixing angle differences are really sensitive to the signs of quark masses.
 The case of the KM phase difference is more subtle, due to the contribution from both charm and top quarks.

\begin{figure}
\includegraphics[scale=0.5]{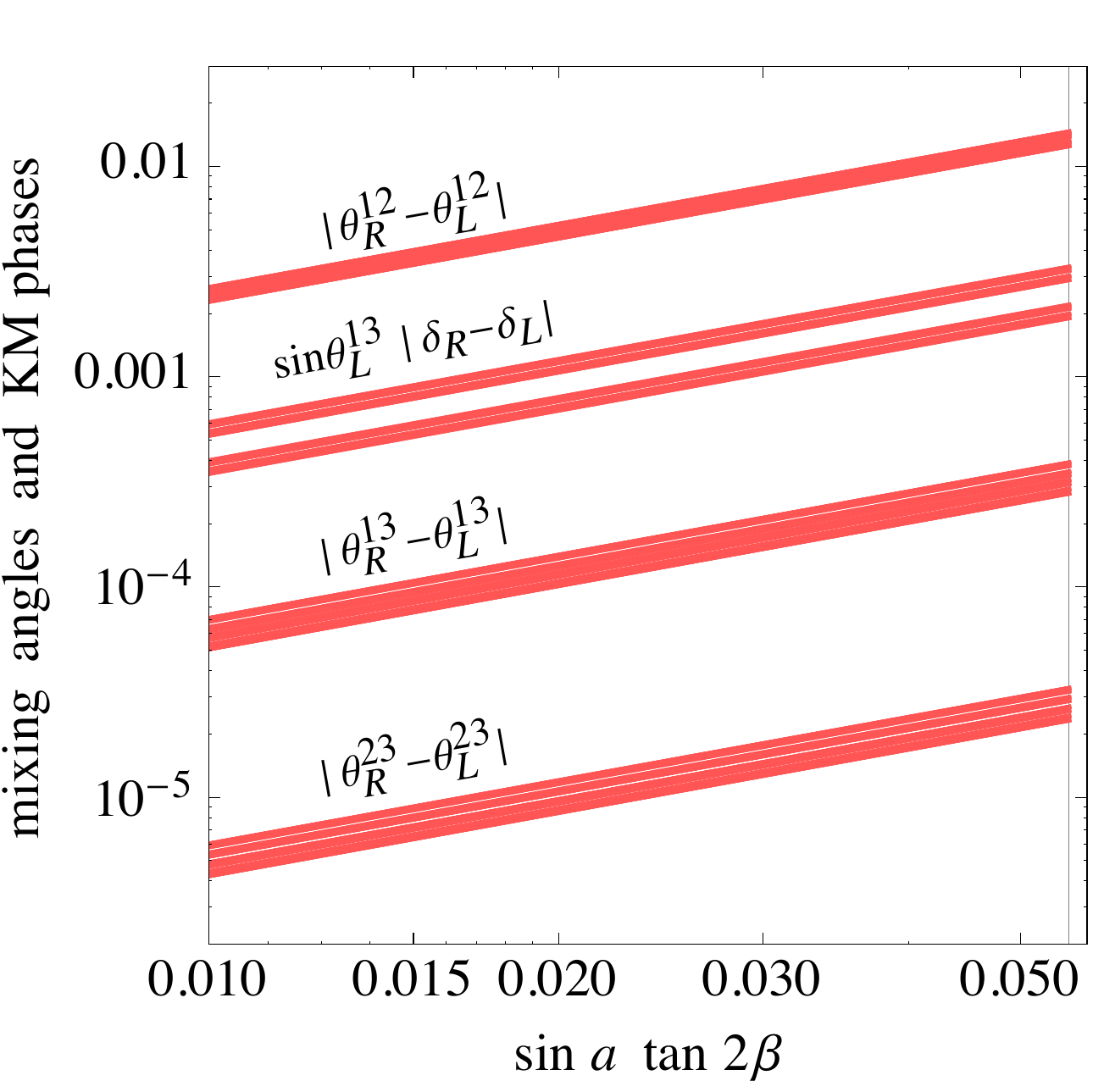}
\caption{First order of the absolute value of right handed mixing angles and KM phases differences.  All signs are included and the stability of the result is manifest.   }
\label{fig:VR-angles-signs}
\end{figure}

  In Fig.~\ref{fig:VR-angles-signs} we plot the impact of the quark mass signs on the absolute values of mixing angle and KM phase differences computed at the first order in $s_a  t_{2\beta}$. We include all the possible sign variations, which results in band like spreads of these quantities. The stability of mixing angle differences is remarkable; for the KM phases there is some dependence on the signs manifested in two smaller bands,  very near to each other.
  When discussing these quantities phenomenologically one can safely ignore different signs, at least at  first order in $s_a  t_{2\beta}$.

As can be seen in \eqref{eq:extphasesw1}-\eqref{eq:extphasesw5}, the external phases depend quite sensitively on the signs of quark masses, except for $\omega_2$ and $\omega_3$.  This is to be expected, for sign changes are equivalent to phase changes from zero to $\pi$. One has to live with this aspect of the theory when discussing the CP violating processes and it must be taken into account when setting limits on the LR scale.

\subsection{Higher orders: external phases}

As one can see from Fig.~\ref{fig:VR-phases},  first order expressions for the 
 external phases \eqref{eq:extphasesw1}-\eqref{eq:extphasesw5} are a good approximation only for small $s_at_{2\beta}$.  We now address the question of the convergence of the external phases as one approaches to the upper bound of   $s_at_{2\beta}$.

Using the standard parametrization for $V_R$,  we find the higher orders of the phase $\omega_3$ (other phases can be treated in a similar way). For this, it is enough to focus on  the 3-3 element of $V_R$
\begin{equation}\label{eq:VR33}
(V_R)_{33}=e^{i \omega_3} \cos \theta_R^{23}\cos\theta_{R}^{13}
\end{equation}

To get  $\omega_3$,   we first expand $(V_R)_{33}$ up to third order using the expressions given in the Appendix~\ref{HigherO}. Keeping only 
dominant terms up to  third order,   one obtains 
\begin{align}
\begin{split}
&(V_R)_{33}\simeq c_{23}c_{13} \bigg[ 1-i s_at_{2\beta}\frac{m_t}{2m_b}\big(1+s_{23}^2+s_{13}^2
\big)
\\
&
\!-\frac{1}{2}\bigg(\! s_at_{2\beta}\frac{m_t}{2m_b}\!\bigg)^{\!\!2}\!\!\big(1\!+\!s_{23}^2\!+\!s_{13}^2
\big)^2
%
\!\!-i \bigg(\! s_at_{2\beta}\frac{m_t}{2m_b}\!\bigg)^{\!\!3}\!   \frac{m_s}{m_b}s_{23}^2
\bigg]
\end{split}
\end{align}
We can see that, with the exception of the first, the odd terms in the expansion are negligible. Thus,
summing the series, one  has to a very good approximation
\begin{align}\label{eq:VR33sum}
&(V_R)_{33}\!\simeq \!c_{23}c_{13}  \exp\!\bigg\{ \!\!- \! i \arcsin \bigg[\!s_at_{2\beta}\frac{m_t}{2m_b} \big(1\!+\!s_{23}^2\!+\!s_{13}^2
\big) \!  \bigg]    \bigg\} 
\end{align}
From \eqref{eq:VR33} and  \eqref{eq:VR33sum} one gets  
\begin{align}\label{eq:w3summed}
\sin \omega_3&  \simeq -s_at_{2\beta}\frac{m_t}{2m_b}(1+s_{23}^2+s_{13}^2
\big) 
\end{align}
The existence  of the solution is then guaranteed by \eqref{eq:limit}.  

To get the rest of the phases, it turns out to be sufficient to substitute $\omega_3$ in\eqref{eq:extphasesw1}-\eqref{eq:extphasesw5} by the corrected one given above.
 This takes into account  only  higher order corrections to the  dominant  $m_t/m_b$ term; corrections to the rest of the factors is straightforward but  we have not included them due to their negligible effect.
\begin{figure} 
  \hspace{-0.3cm}  \includegraphics[scale=0.515]{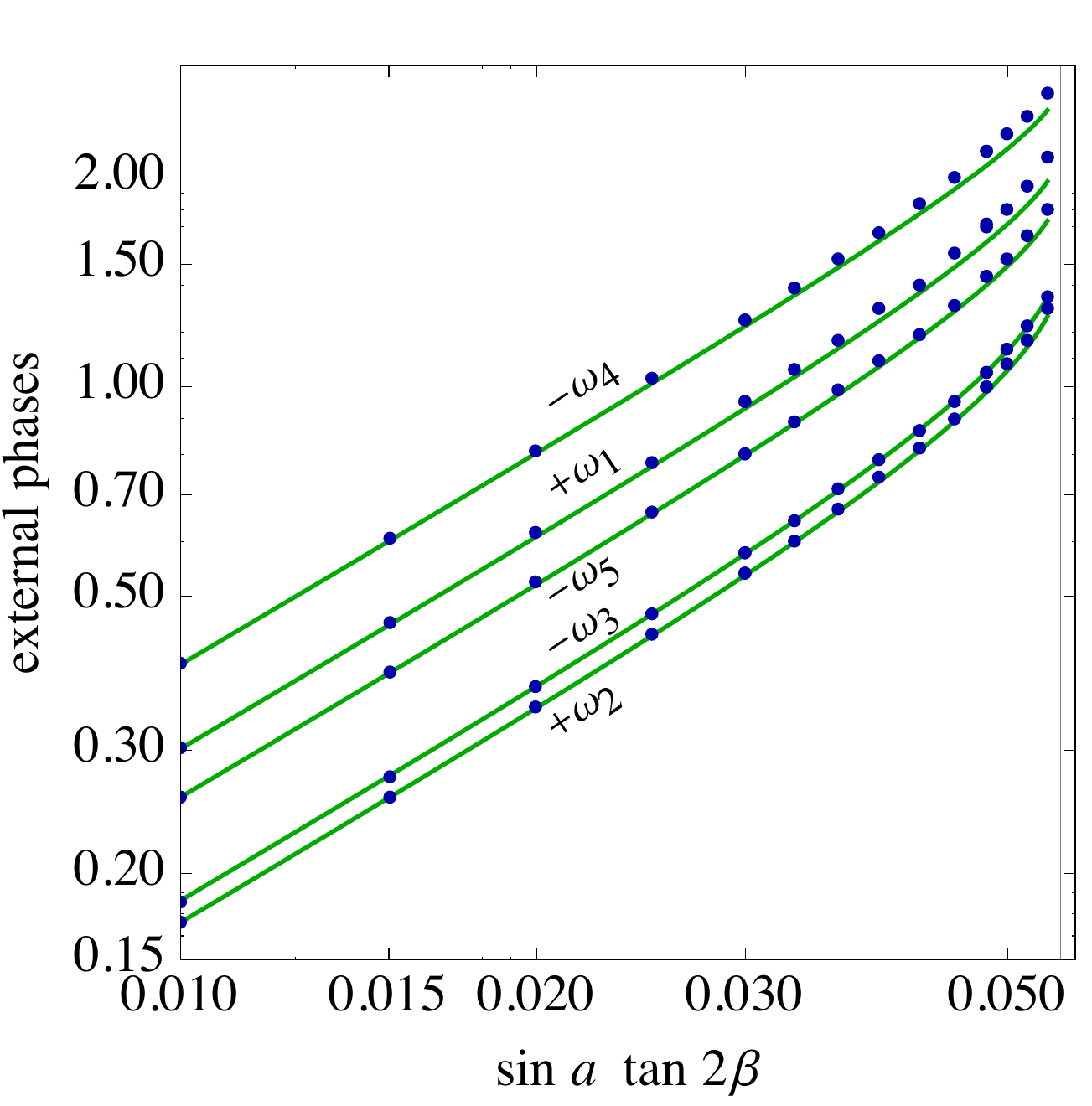} 
\caption{External phases with $\sin \omega_3= -s_at_{2\beta}m_t/(2m_b)$  corresponding to higher order expression (green line) and  the  numerical solution of the exact equation (blue dots). 
The  agreement with the numerical solution is excellent and covers the whole parameter space.}
\label{fig:VR-phases-b}
\end{figure}
The final result is  shown in Fig.~\ref{fig:VR-phases-b}. Notice that the phase $\omega_3$ controls the range of validity of the whole solution of $V_R$, and it stops at $\sim\pm\pi/2$ when $s_at_{2\beta}$ reaches its upper bound \eqref{eq:limit}.  The agreement with the numerical results is now manifest in the whole parameter space. Once again, we confirm that the perturbation expansion in the small $s_at_{2\beta}$ works well already at first order and the convergence of the external phases is  under control.

  Last but not least, for illustrative purposes, the reader is referred to Appendix~\ref{0mixing} for the exact solution of the RH phases in the non-realistic case of vanishing mixing angles.

\section{Phenomenological applications }
\label{pheno}

 \subsection{RH charged gauge boson at colliders}
 
  The physics of the $W_R$ charged gauge boson needs the knowledge of $V_R$ as much as one needs to know the CKM matrix in the case $W$. The proximity of LH and RH mixing angles tells us that the $W_R$ production at the hadronic colliders  proceeds basically  in  the same manner as the production of the $W$ boson. This is useful since both CMS and ATLAS~\cite{Khachatryan:2014dka, Aad:2014aqa} in most of their searches for a generic $W'$ assume  identical production rates as for the $W$ boson. 
  
    More precisely, the production strength of $W_R$ is proportional to 
 \begin{align}\label{eq:VR11}
|(V_R)_{11}|\simeq  c_{12}\bigg(1+ s_at_{2\beta} \frac{m_t}{m_s}   s_{12}  s_{23} s_{13} s_{\delta} \bigg)
\end{align}
      It is easy to see that  $|(V_R)_{11}|^2$ differs from $|(V_L)_{11}|^2$ by at most  a percent; thus to a great precision the production rate of $W_R$ is equal to that of $W_L$. Similarly, the decay rates into top and bottom, or di-jets in general are to an excellent approximation given by assuming the same left and right couplings. Needless to say, it is possible to have LR symmetry broken at large scales; in such a case  one has to run down our expressions to the relevant energies. The predictions will then be modified  in a calculable manner. 
      
        On the other hand, the leptonic decay rates of LH and RH gauge bosons are completely different in general, due to the seesaw mechanism behind neutrino mass. In this case 
         $W_R$ can decay into two jets and two same sign charged leptons \cite{Keung:1983uu}, an appealing possibility of  direct observation   of  lepton number violation. 
        
\subsection{Neutrinoless double beta decay}

 The LR symmetric model offers naturally a new contribution~\cite{Mohapatra:1979ia} to the neutrinoless double beta decay ($0\nu\beta\beta$) through the exchange of $W_R$ and RH neutrinos, in complete analogy with the usual contribution of Majorana light neutrinos.  
      It goes through the RH gauge currents and RH neutrino mass, depending crucially on both lepton and quark RH mixing matrices. The total $0\nu\beta\beta$ rate can then be described by the effective mass parameter  \cite{Tello:2010am}
     \begin{equation}
     \!\!\!\! |M_{\nu+N}^{ee}|\!=\! \sqrt{|(M_{\nu})_{ee}|^2+ \frac{|(V_R)_{11}|^4}{|(V_L)_{11}|^4}
     \frac{M_{W_L}^8}{M_{W_R}^8}\left|\,\left( \frac{k^2}{ M_N }\right)_{\!\!ee}\right|^2 }
     \end{equation}
     where $k$ is a measure of the neutrino virtuality, $M_{\nu}$ and $M_{N}$ are the mass matrices of light and heavy neutrinos, respectively. As we mention above, the lepton mixing is not predicted by the theory, and its determination would require the eventual direct observation of lepton number violation along the lines of the KS process. However, it is reassuring to know that the RH quark vertex $(V_R)_{11}$, as shown in  \eqref{eq:VR11},  can be predicted with arbitrary precision, in the same manner as in the case of the $W_R$ production at hadron colliders.  
     
       An additional contribution to $0\nu\beta\beta$ mediated by the RH doubly charged scalar is disfavoured by lepton flavor violating constraints \cite{Tello:2010am}. The rate can also proceed through the Dirac masses of neutrinos but, for RH neutrinos visible at the LHC, it  amounts to a sub-leading effect  \cite{Nemevsek:2012iq}. Nevertheless, these sub-leading contributions also depend on the quark mixing $V_R$ and their precise evaluation is thus equally feasible.

\subsection{The theoretical limits from K and B physics}

Another nice example of the relevance of $V_R$ are the low energy limits on the LR scale from the  kaon and B-meson physics. They depend crucially on the RH mixings, and we will illustrate it on a few examples.

{\bf K meson physics.} Let us discuss the $K- \bar K$ mass difference. As is well known, in this case the dominant effect comes from the charm quark in the box diagram with the $W_L$ and $W_R$ charged gauge bosons. Only the real part of the amplitude contributes and it is proportional to the following form of the RH mixing matrix 
\begin{align}\label{eq:KLKS-VRterm}
\begin{split}
&\hspace{-0.25cm}\text{Re}[\mathcal{A}_{LR}(K\rightarrow \bar{K})]\!\propto\!\text{Re}\,[ (V_L)_{21}(V_L^*)_{22}(V_R)_{21}(V_R^*)_{22}]\\
&\hspace{-0.24cm}\simeq s_{d}s_{s} c_{12}^2s_{12}^2
\\
&\hspace{-0.24cm}\times \bigg[1-
\frac{1}{8} \big(s_at_{2\beta} \big)^2 \bigg( \frac{c_{12}^2}{m_s}+\frac{s_{12}^2}{m_d}\bigg)^{\!2} \big(m_c c_{23}^2+m_t s_{23}^2\big)^2 \bigg] 
\end{split}
\end{align}
where we have used the   formula   for $V_R$  up to second order (for higher orders  see Appendix~\ref{HigherO}).
Alternatively, neglecting the small difference  between left and right mixing angles 
 one gets  $c_{12}^2 s_{12}^2\,\!\cos (\omega_4\! -\! \omega_5) $.
Using \eqref{eq:extphasesw4} and \eqref{eq:extphasesw5} one arrives again at \eqref{eq:KLKS-VRterm}.
Due to different sign options in $V_R$, the cosine can be either $-1$  (and remain stable for any $s_a t_{2\beta}$ because of the cancellation of the first and second term in the rest of the difference  $\omega_4-\omega_5$) or lie between $1$ and $\sim0.7$, in this case depending also on the value of $s_a t_{2\beta}$.

 Notice an interesting fact that helps the stability of the limit on the $W_R$ mass. The difference $\omega_4 - \omega_5$
 is substantially smaller that the individual $\omega_4$ and $\omega_5$ phases as seen from 
 \eqref{eq:extphasesw4}-\eqref{eq:extphasesw5},
 and we can see clearly how the limit on the scale actually emerges. 
 
  In a similar manner one can compute the imaginary part of the amplitude, which we leave as an exercise to the reader.
  
{\bf B meson physics.} In this case the relevant quantity to compute to the leading order, due to the top quark dominance, is
(see e.g.~\cite{Maiezza:2010ic})
 \begin{equation}\label{bmeson}
r_{d_i} \propto (V_R^*)_{33}(V_R)_{3i}
\end{equation}
where $d_{i}$ stands for strange ($i=2$) or down  ($i=1$) quark.
Neglecting the tiny difference between the left and right mixing angles,  one gets the leading contribution
\begin{align}\label{bmeson}
r_{d}& \propto   c_{23}c_{13}
\left(
s_{12}s_{23}
-c_{12}c_{23}s_{13}e^{i\delta_R}
\right)
e^{i\omega_4 } \\[3pt]
r_{s} &\propto - c_{23}c_{13} c_{12}s_{23}e^{i\omega_5}
\end{align}
with  $\delta_R$, $\omega_4$ and $\omega_5$ given approximately by \eqref{eq:Delta}, \eqref{eq:extphasesw4} and \eqref{eq:extphasesw5}, respectively at first order  (to account for  higher orders corrections one should use also \eqref{eq:w3summed}).
Again these analytical results  facilitate numerical studies.

\subsection{The physics of the heavy scalar doublet from the bi-doublet}

       It has been known for a long time that the second doublet in the bi-doublet, due to the large amount of flavor violation it induces in neutral currents,   has to have a large mass on the order 10-15 TeV or so (see for example~\cite{Maiezza:2010ic},~\cite{Blanke:2011ry}). This is not a problem since it gets its 
       mass~\cite{Senjanovic:1979cta} from the large triplet vev, responsible for the masses of heavy gauge boson. Its interactions can be deduced from the Yukawa in \eqref{eq:quarks&phi}
\begin{align}\label{eq:quarks&phi2}
\begin{split}
- L_Y=&  - \overline{q_{L}}\bigg[  \frac{M_u}{v}h  -\frac{M_{d}+e^{-ia}  s_{2\beta}M_u}{v c_{2\beta}}H  \bigg]u_R 
\\[5pt]
  +\overline{q_{L}}&\bigg[ \frac{M_{d}}{v} i \sigma_2h^*    
-\frac{ M_u+e^{ia}s_{2\beta}M_{d}}{vc_{2\beta}} i \sigma_2H^*  
\bigg]d_R+\text{h.c.}
\end{split} 
\end{align}
where  
\begin{equation}\label{eq:physicalphi}
h= c_{\beta}\phi_1+e^{-i a}s_{\beta}\phi_2 ,\quad  H=-e^{i a}s_{\beta}\phi_1+c_{\beta}\phi_2
\end{equation}
$h$ and $H$  are the doublets with and without a vev, respectively, and $\phi_1$ and $\phi_2$ are SU(2)  doublets with $Y=-1$, forming the  bi-doublet $\Phi=(\phi_1,i\sigma_2\phi_2^*)$. The light doublet, $h$, is effectively the SM one, while $H$ stands for the heavy doublet. Since $H$ has to weigh more than 10 TeV, they can be taken with an excellent precision to be mass eigenstates. The Yukawa interaction of $H$ in the physical basis becomes
\begin{align}
\begin{split}
- L_{H} & =  H^0  \overline{u_L} \, \dfrac{V_Lm_{d}V_R^{\dagger}+e^{-ia}s_{2\beta}m_u}{v c_{2\beta} } u_R 
  \\[3pt]  
  & + H^{0*}  \overline{d_L}\, \dfrac{V_L^{\dagger} m_uV_R+e^{ ia}s_{2\beta}m_{d}}{v c_{2\beta} } d_R 
  \\[3pt]
    &+ H^-  \overline{d_L}  \,\dfrac{m_{d}V_R^{\dagger}+e^{-ia}s_{2\beta}V_L^{\dagger} m_u}{v c_{2\beta} } u_R 
    \\[3pt]
   & -   H^+  \overline{u_L}\, \dfrac{m_u V_R 
  + e^{ ia}s_{2\beta}V_L m_{d}}{v c_{2\beta} }    d_R+\text{h.c}
  \end{split}
\end{align}

   The above interactions depend crucially on $V_R$ just like the gauge interactions of the heavy gauge bosons. This is relevant  for the low energy processes such as $K_L-K_S$ mass difference and for the production and identification of the heavy doublet $H$ at future hadronic colliders. 
   
Regarding $K-\bar{K}$ mass difference. The neutral component of the heavy doublet  adds coherently to the box diagram and will not affect the stability of the bound on the  $W_R$ mass.  The tree level amplitude is proportional to  
 \begin{align}
\text{Re}[\mathcal{A}_{H}(K\rightarrow \bar{K})]\propto \text{Re}[(V_L^{\dagger} m_u V_R)_{21} (V_R^{\dagger} m_u V_L)_{21}]
\end{align}
which, due to gauge invariance~\cite{Basecq:1985cr}, has the same flavor  dependence as in the usual $W_L - W_R$ box diagram.   
Using as before higher orders of $V_R$ one gets the same leading term as in \eqref{eq:KLKS-VRterm}.

   Similarly,  the production and decays of the heavy doublet depend strongly on the quark mixings $V_R$ and,  for small values of $\beta$, it  may  dominate the interaction. 
   Notice an important fact: $H^0$ decays principally in $b \bar b$, a very clear signature. The other decay rates, such as say, $t \bar t$ depend on $\beta$.

\subsection{The strong CP parameter}

     As argued originally in~\cite{Beg:1978mt}, generalized parity guarantees the vanishing of the QCD strong CP parameter $\theta$ before the symmetry breaking. This makes the effective $\bar \theta$ calculable through the knowledge of $V_R$, as done recently approximately in~\cite{Maiezza:2014ala}. As a particularly nice illustration of the power of our results,  in what follows we  calculate  exactly  the first order of the series expansion of  $\bar \theta$ in terms of the small parameter $s_a t_{2\beta}$. 
     
     In terms of the matrices that diagonalize the up and down quark mass matrices, one has 
 \begin{equation}\label{def-theta-bar}
     \bar\theta = \text{arg det }\, U_L^\dagger  U_R \, D_L^\dagger D_R
     \end{equation}
    Now defining, as in Appendix \ref{LO},  $ U_u=   U_L^{\dagger}U_R  , \quad  U_d=   D_L^{\dagger}D_R$, and using  $V_L U_d=U_u  V_R$ 
     one gets  (noting also that $\text{arg det } V_L=0$)
 \begin{equation}
     \bar{\theta} = \text{arg det } U_u^2 V_R 
    \end{equation}  
    Using the explicit form of $U_u$  as a matrix square root from \eqref{eq:Uu}, it gives
    \begin{equation} \label{eq:exactThetaBar}
     \bar{\theta}=\text{arg det}\left[ V_R  + i s_a t_{2\beta}\left(  t_\beta e^{-i a} V_R
   +   m_u^{-1}  V_L m_d\right) \right] 
     \end{equation}
     This is an  tree-level expression, depending mainly on quark masses and mixing matrices.   It could serve to compute $\bar \theta$ with an arbitrary precision.
     We can expand \eqref{eq:exactThetaBar} in the small parameter $s_a t_{2\beta}$
     and obtain the  leading term\footnote{Using the expansion  $\text{arg det} \left[ 1 + i \epsilon M\right] = \epsilon{\text{Re}} (\text{tr}M )+O( \epsilon^2 )$.
     }  
     \begin{align}\label{eq:theta}
\bar{\theta}&= s_a t_{2\beta}  \,\frac{1}{2}  \, \text{Re} \,\text{tr} \bigg( m_u^{-1}V_Lm_d  V_L^{\dagger}   - m_d^{-1}V_L^{\dagger}m_u  V_L  
\bigg) 
\end{align}

     This expression and the exact one above, disagrees with the approximate form presented in~\cite{Maiezza:2014ala}. The dominant term, though, is the same and is given by
    $ \bar\theta \simeq s_a t_{2\beta} m_t/2 m_b $, and could have been readily guessed by ignoring the small third generation CKM mixing angles.
      It is 
    obviously huge unless $s_a t_{2\beta}$ is vanishingly small. This forces a large bound on the LR scale as argued in~\cite{Maiezza:2014ala}.
    
    How solid is this argument? In order to answer this, let us digress for a moment and discuss the question of strong CP in the SM. The fact that, experimentally, the strong CP parameter must be small $\bar \theta \lesssim 10^{-10}$ has been coined the strong CP problem. But is this  really a problem? In other words, is the perturbative contribution to $\bar \theta$ much bigger than the experimental upper bound? The answer is no, in fact it is much smaller as we now remind the reader.
    
    It is true that the SM does not predict the value of $\bar \theta$ and it is a puzzle why it is so small, in view of the fact that the weak CP violation parameter is much bigger. It is also true that $\bar \theta$ appears to be divergent in perturbation theory, however, not before the sixth loop~\cite{Ellis:1978hq} and for any reasonable value of the cut-off it is negligibly small. For example, for $\Lambda_{cutoff} \simeq M_{Planck}$ one gets $\bar \theta \simeq 10^{-19}$~\cite{Ellis:1978hq}, orders of magnitude below the experimental limit. In other words, the question of the strong CP violation is a question of high energy physics, not a question of the SM itself. More precisely, the large scale that sets the value of $\bar \theta$ decouples effectively from low energy processes, in the same way  the scale of grand unification decouples from low energies, except from fixing the weak mixing angle~\cite{Georgi:1974yf} (of course, there is the new possibility of inducing a proton decay, but this is consistent with the decoupling in usual interactions).
       
         A nice example of new high energy physics is the Peccei-Quinn (PQ) symmetry~\cite{Peccei:1977hh}. It fixes $\bar \theta$ no matter how large its scale is, and for practical purposes it decouples from low energy phenomena, the exception being a highly sensitive physics of the axion, just as proton decay in grand unification. The fact that the PQ symmetry does the job in the LR theory without affecting low energy physics has been emphasised already in \cite{Maiezza:2014ala}. The message from this is simple: one should not worry about $\bar \theta$ in theories at low, or relatively low, energies, such as the SM or the LR symmetric theory. This is a philosophy we will succumb to, and simply imagine a new physics such as PQ symmetry that takes care of the strong CP issue. In other words, we will allow $s_a t_{2 \beta}$ to take any value below its strict phenomenological bound in \eqref{eq:limit}. Independently of how this is settled, and independently of the precise value of $\bar \theta$ one could obtain, however, a lower limit on the LR scale by considering the electric dipole moments of atoms, such as mercury. Although well measured, its theoretical value is plagued by large uncertainties, so setting a reliable limit will have to wait some more.

\section{ Summary and outlook }
\label{s&o}
  In a recent Letter we were able to elucidate the long awaited form of the RH quark mixing matrix $V_R$ in the minimal LR symmetric model augmented with generalized parity. We found exact equations, valid in the entire parameter space, which allow for its numerical determination. We gave there an approximate form for $V_R$ as a leading term in the expansion of  a small parameter which measures the departure from the hermiticity of the quark mass matrices. Moreover, we argued in favor of the proximity between left and right mixing angles, using the important fact that in the two generation case $\theta_R=\theta_L$. The small CKM mixing angles, together with a small ratio between strange and bottom quark masses,  then guarantee practically equal mixing angles in the realistic three generation case. 
    
    In this companion paper we provided a more detailed and complete discussion, paying special attention to the convergence of the employed expansion by calculating the second and third order of the series. 
    In the case of mixing angles, we find an excellent agreement with the numerical results already at the first order, while in the case of the phases 
     in general one needs higher orders. 
    
     We also discussed here a number of phenomenological applications that depend strongly on the form of the RH quark mixing.  Our study shows the importance of the explicit analytical and numerical knowledge of $V_R$. In the case of  $W_R$ we demonstrate that the collider production and signature in di-jet decays is perfectly well described with the usual SM couplings of the $W$ boson, except the opposite chirality. Similarly,  we determined the strength  of the right-handed gauge currents which control the rates of neutrinoless double beta decay.
      We also show how the decay rates of the heavy scalar doublet in the bi-doublet can be studied with great precision now that $V_R$ is made transparent.
        As a nice example of the application of our results we give an exact and a leading term expression for the strong CP parameter $\bar \theta$. More important, we gave an explicit expression for the dependence of $K_L-K_S$ mass difference on $V_R$ and showed that the stability of the resulting limit on the $W_R$ mass is due to a partial cancellation of external phases.
      
       One last comment. The determination of $V_R$, as we showed here, requires only the minimal quark Yukawa sector and it does not depend, at least not at the tree level, on the situation in the leptonic sector or on  the details of the parity breaking at the high scale. In other words, it does need the Majorana seesaw picture of neutrinos and could as well apply to the original LR model~\cite{lrmodel}. After all, neutrinos could be Dirac particles~\cite{Branco:1978bz}, no matter how appealing we find the seesaw picture. All that matters here is that our findings hold true in any LR symmetric model with the single bi-doublet $\Phi$. While we prefer the seesaw picture which makes the LR theory complete and predictive when it comes to neutrino mass, it is up to experiment to decide.

\subsection*{Acknowledgments}

 We are grateful to Federica Agostini, Alejandra Melfo, Darius Faroughy, Fabrizio Nesti, Juan Carlos Vasquez and Yue Zhang for numerous discussions and comments, and for careful reading of the manuscript. 
 

\onecolumngrid

\appendix

\section{Leading order}
\label{LO}
In order to determine $V_R$ one has to deal with a matrix equation involving  square roots of matrices \cite{history}. 
For simplicity and illustration we begin  with the following square root
\begin{equation}\label{eq:example}
\sqrt{m^2+ i \epsilon A}
\end{equation}
where $m$ is a diagonal matrix, $A$ a matrix, and $\epsilon$ an  expansion parameter. The problem is to express the above square root in terms of the elements of $m$ and $A$ for small $\epsilon$. There are different ways of attacking this problem; we will work in 2 dimensions, where a root of matrix itself has a simple analytical form. The resulting expression turns out to be valid for any number of dimension.   

Using the formula for square root in the $2\times2$ case we have (taking all signs positive to ease the presentation)
\begin{equation}\label{eq:example2x2}
 \sqrt{m^2+ i \epsilon A}=\frac{  (m^2+ i\epsilon A) + \sqrt{d} I_{2\times2} }{\sqrt{t +2 \sqrt{d}}}  
\end{equation}
where $d=\det (m^2+ i \epsilon A)$ and $t=\text{tr}(m^2+ i \epsilon A)$. Expanding  first
\begin{align}
&\sqrt{d}=m_1m_2 \bigg[ 1+i \,\frac{\epsilon}{2} \, \text{tr} (m^{-2}A) \bigg] \\
&\frac{1}{\sqrt{t +2 \sqrt{d}  }}= \frac{1}{m_1+m_2}\bigg(1 -   \frac{i\epsilon}{2} \frac{\text{tr}A+m_1 m_2\text{tr} (m^{-2}A) }{(m_1+m_2)^2}\bigg)
\end{align}
and using them in \eqref{eq:example2x2} one arrives at
\begin{align}\label{eq:rootexpansion}
 \left(\sqrt{m^2+ i \epsilon A}\,\right)_{ij}&=m_i \delta_{ij}+i\epsilon\frac{ A_{ij}}{m_i+m_j} +O (\epsilon^2)
\end{align}

It is easy to check that this expansion is valid  regardless the number of dimensions. 
Still, here we outline the general steps needed to find the square  root of a matrix $M$ of arbitrary dimensions. This is useful also when solving the equation numerically.
 The first step is to decompose  $M$ into
\begin{equation}
M=XJX^{-1}
\end{equation}
where $J$ is called the normal form of $M$. Once $J$ and $X$ are found (see \cite{Gantmacher} for the details), one finds  the root as
\begin{equation}
\sqrt{M}=XY\sqrt{J}Y^{-1}X^{-1}
\end{equation}
Here, the matrix $Y$ parametrizes a set of continuous  solutions and it is only present in some very special cases.  

Coming back to 
 the square root in \eqref{eq:example}, now working in any number of dimensions, 
  the above decomposition  allows a straightforward  expansion in terms of the small parameter. 
 The result being once again formula \eqref{eq:rootexpansion}.

In what follows we outline the  steps needed to determine $V_R$. It will be  useful to introduce unitary matrices $U_u$ and $U_d$  which become diagonal sign matrices  when the corresponding  mass matrices are hermitian. In the notation of \eqref{notationUD}
   \begin{equation} \label{UuUd}
   U_u=   U_L^{\dagger}U_R  , \quad  U_d=   D_L^{\dagger}D_R
 \end{equation}
Then from \eqref{relationsMuMd-1} and \eqref{relationsMuMd-2} one finds 
 \begin{align}\label{eq:Uu}
&\!\!\!U_u=\frac{1}{m_u}\sqrt{m_u^2+is_ a t_{2\beta}   \left(t_{\beta}e^{-i a}m_u^2+m_uV_L m_d V_R^{\dagger}\right) } \\[3pt]
 \label{eq:Ud}
&\!\!\! U_d=\frac{1}{m_d}\sqrt{m_d^2- i s_ a t_{2\beta}  \left(t_{\beta}e^{ia}m_d^2+m_dV_L^{\dagger}m_uV_R\right) }
\end{align}
 One has an additional relation which arises from the definition of the mixing matrices
\begin{equation} \label{eq:VR}
V_L U_d=U_u  V_R
\end{equation} 
 Together with \eqref{eq:Uu} and \eqref{eq:Ud} it allows the determination of $V_R$ in terms of $V_L,m_u,m_d,a$ and  $\beta$. .
Indeed using  \eqref{eq:rootexpansion}
 one can expand the square roots in $U_u$ and $U_d$  in \eqref{eq:Uu}  and \eqref{eq:Ud} in powers of $s_at_{2\beta}$ 
\begin{align}\label{eq:Uuexpanded}
(U_u)_{ij}&= (S_u)_{ij}+i    s_a t_{2\beta}\left( \frac{ t_{\beta }}{2}(S_u)_{ij}+\frac{( V_L m_d V_R^{\dagger})_{ij}}{\hat{m}_{u_i}+\hat{m}_{u_j}}\right)   +O( s_a^2 t_{2\beta}^2 )\\
\label{eq:Udexpanded}
(U_d)_{ij}&=(S_d)_{ij}-i  s_a t_{2\beta} \left(\frac{ t_{\beta }}{2}(S_d)_{ij}+  \frac{( V_L^{\dagger}m_uV_R)_{ij} }{\hat{m}_{d_i}+\hat{m}_{d_j}}  \right)
+O( s_a^2 t_{2\beta}^2 )
\end{align}
where $S_{q_i}=\text{diag}(s_{q_i})$, $\hat{m}_{q_i}=s_{q_i}m_{q_i}$ and $s_{q_i}$ are $\pm$ signs.
Using these expressions  in \eqref{eq:VR} one can find $V_R$ as a  power series in $s_at_{2\beta}$
\begin{equation}
V_R=V_R^{(0)}-i s_a t_{2\beta }V_R^{(1)}+\cdots
\end{equation}
The explicit form of $V_R^{(0)}$ and $V_R^{(1)}$ can be read-off from  \eqref{eq:master}, with $V_R^{(0)}=V_L$ up to  sign matrices, and $V_L^{(1)}$ the term in brackets.
There are $2^{(2n-1)}$ independent solutions for $n$ generations, due to the square root nature of \eqref{eq:Uu} and \eqref{eq:Ud}. The rest is found through $V_L\rightarrow S_uV_L S_d$ and $m_{q_i}\rightarrow s_{q_i} m_{q_i}$, where $S_u=\text{diag}(s_{u_i})$, $S_d=\text{diag}(s_{d_i})$ with, as before, $s_{q_i}$ being $\pm$ signs.

\section{Higher order terms }
\label{HigherO}
 
 Let us turn again to the two-dimensional prototype case.  From  \eqref{eq:example2x2} it is straightforward to get higher order terms

\begin{align}\label{third-order}
\left(\frac{1}{m}\sqrt{m^2+ i\epsilon m A}\right)_{ij}&= \delta_{ij}+i \epsilon \frac{A_{ij}}{m_i+m_j}+\epsilon^2  \frac{A_{ik}m_kA_{kj}}{(m_i+m_k)(m_k+m_j)(m_i+m_j)}  \nonumber \\
& -i \epsilon^3  \frac{A_{ik_1}m_{k_1}A_{k_1k_2}m_{k_2}A_{k_2 j}}{(m_i+m_{k_1})(m_{k_1}+m_{k_2})(m_{k_2}+m_j)(m_i+m_j)}\left(\frac{1}{m_i+m_{k_2}}+  \frac{1}{m_j+m_{k_1}} \right)+O(\epsilon^4)
\end{align}
This can be shown to holds in general for matrices of any dimension.

We can now proceed to determination of the second and third order  of $V_R$ in terms of $s_at_{2\beta}$. For the sake of space and clarity, we present only  contributions of the type $m_u/m_d$ due to the large   top to bottom quark masses ratio, since these are the only ones that can compensate for the smallness of $s_at_{2\beta}$. As we have seen already at the first order, all other terms are sub-leading at this level. For this purpose it is enough we take  $U_u\simeq S_u$ and compute only   higher orders of  $U_d$.

Using \eqref{third-order} one has up to the third order
%
\begin{align}\label{eq:Uexpanded2}
(U_d)_{ij}&\simeq (S_d)_{ij}-i  s_a t_{2\beta} \frac{( V_L^{\dagger}m_uV_R)_{ij} }{\hat{m}_{d_i}+\hat{m}_{d_j}}  + s_a^2 t_{2\beta}^2  \frac{( V_L^{\dagger} m_u V_R)_{ik}(m_d V_L^{\dagger} m_u V_R)_{kj}}{(\hat{m}_{d_i}+\hat{m}_{d_k})(\hat{m}_{d_k}+\hat{m}_{d_j})(\hat{m}_{d_i}+\hat{m}_{d_j})}
\nonumber \\
&+ is_a^3 t_{2\beta}^3 \frac{( V_L^{\dagger} m_u V_R)_{ik_1} (m_d V_L^{\dagger} m_u V_R)_{k_1k_2}( m_dV_L^{\dagger} m_u V_R)_{k_2j}}{(\hat{m}_{d_i}+\hat{m}_{d_{k_1}})(\hat{m}_{d_{k_1}}+\hat{m}_{d_{k_2}})(\hat{m}_{d_{k_2}}+\hat{m}_{d_j})(\hat{m}_{d_i}+\hat{m}_{d_j})}\left(\frac{1}{\hat{m}_{d_i}+\hat{m}_{d_{k_2}}}+\frac{1}{\hat{m}_{d_j}+\hat{m}_{d_{k_1}}}\right)
\end{align}
%
where $S_{d}=\text{diag}(s_{d_i})$, $\hat{m}_{d_i}=s_{d_i}m_{d_i}$ and $s_{d_i}$ are $\pm$ signs. Expressing also $V_R$  as a series of powers of $s_at_{2\beta}$  
\begin{equation}
V_R=V_R^{(0)}-i s_a t_{2\beta }V_R^{(1)}-(s_at_{2\beta})^2V_R^{(2)}+i (s_a t_{2\beta })^3 V_R^{(3)}+O(s_a^4t_{2\beta}^4)
\end{equation}
and using it together with  \eqref{eq:Uexpanded2} in \eqref{eq:VR} one gets, after equating  each order of $s_at_{2\beta}$,   the following recurrence relations
%
  \begin{align}
 (V_R^{(0)})_{ij}&=(S_u V_L S_d)_{ij}\\
  (V_R^{(1)})_{ij}&\simeq(S_u V_L)_{ik} \frac{( V_L^{\dagger}m_uV_R^{(0)})_{kj} }{\hat{m}_{d_k}+\hat{m}_{d_j}}
  \\
   (V_R^{(2)})_{ij}&\simeq (S_u V_L)_{ik_1} \left(   \frac{( V_L^{\dagger}m_uV_R^{(1)})_{k_1j} }{\hat{m}_{d_{k_1}}+\hat{m}_{d_j}}     
-
\frac{( V_L^{\dagger} m_u V_R^{(0)})_{k_1k_2}(m_d V_L^{\dagger} m_u V_R^{(0)})_{k_2j}}{(\hat{m}_{d_{k_1}}+\hat{m}_{d_{k_2}})(\hat{m}_{d_{k_2}}+\hat{m}_{d_j})(\hat{m}_{d_{k_1}}+\hat{m}_{d_j})} \right)\\
\begin{split}
(V_R^{(3)})_{ij}&\simeq  (S_u V_L)_{i k_1} \bigg[   \frac{( V_L^{\dagger}m_uV_R^{(2)})_{k_1j} }{\hat{m}_{d_{k_1}}+\hat{m}_{d_j}} - 
\frac{( V_L^{\dagger} m_u V_R^{(1)})_{k_1k_2}(m_d V_L^{\dagger} m_u V_R^{(0)})_{k_2j}\!+\!( V_L^{\dagger} m_u V_R^{(0)})_{k_1k_2}(m_d V_L^{\dagger} m_u V_R^{(1)})_{k_2j}}{(\hat{m}_{d_{k_1}}+\hat{m}_{d_{k_2}})(\hat{m}_{d_{k_2}}+\hat{m}_{d_j})(\hat{m}_{d_{k_1}}+\hat{m}_{d_j})}   \\
&
+ \frac{( V_L^{\dagger} m_uV_R^{(0)})_{k_1k_2} (m_d V_L^{\dagger} m_uV_R^{(0)})_{k_2k_3}( m_dV_L^{\dagger} m_u V_R^{(0)})_{k_3j}}{(\hat{m}_{d_{k_1}}+\hat{m}_{d_{k_2}})(\hat{m}_{d_{k_2}}+\hat{m}_{d_{k_3}})(\hat{m}_{d_{k_3}}+\hat{m}_{d_j})(\hat{m}_{d_{k_1}}+\hat{m}_{d_j})}\left(\frac{1}{\hat{m}_{d_{k_1}}+\hat{m}_{d_{k_3}}}+\frac{1}{\hat{m}_{d_j}+\hat{m}_{d_{k_2}}}\right)  \bigg]
\end{split}
  \end{align} 
 %
After solving these equations one arrives at the final result
%
\begin{align} 
&(V_R^{(0)})_{ij}=(V_L)_{ij}\\
&(V_R^{(1)})_{ij}\simeq(V_L)_{ik} \frac{( V_L^{\dagger}m_uV_L)_{kj} }{m_{d_k}+m_{d_j}}\\
&(V_R^{(2)})_{ij}\simeq( V_L)_{ik_1}(V_L^{\dagger}m_uV_L)_{k_1k_2}(V_L^{\dagger}m_uV_L)_{k_2j}\frac{m_{d_{k_1}}}{(m_{d_{k_1}}+m_{d_j})(m_{d_{k_1}}+m_{d_{k_2}})(m_{d_{k_2}}+m_{d_j})} \\
\begin{split}
&(V_R^{(3)})_{ij}\simeq (V_L)_{i k_1} (V_L^{\dagger}m_uV_L)_{k_1k_2}(V_L^{\dagger}m_uV_L)_{k_2k_3}(V_L^{\dagger}m_uV_L)_{k_3j} \\
& \hspace{3cm}\times \frac{m_{d_{k_1}} (m_{d_{k_1}}m_{d_{k_2}}-m_{d_{k_3}}m_{d_{j}})  }{(m_{d_{k_1}}+m_{d_{k_2}})(m_{d_{k_2}}+m_{d_{k_3}})(m_{d_{k_1}}+m_{d_{k_3}})(m_{d_{k_1 }}+m_{d_j}) (m_{d_{k_2}}+m_{d_j})(m_{d_{k_3}}+m_{d_j})}
\end{split}
\end{align}
%
The other solutions can be found by changing, in the above expressions, $V_L\rightarrow S_u V_L S_d$ and $m_{q_i}\rightarrow s_{q_i}m_{q_i}$ with $s_{q_i}$ being $\pm$ signs.  We see once again that a diagonal $V_L$ implies a diagonal for $V_R$, as was found at the first order. This confirms that  small differences between RH and LH mixing angles are protected by small CKM angles. This fact is supported by the exact result of the equality of mixing angles in the two-generation case. We emphasize the parametrization independence of the above expressions, which should simplify the task of applying our results to physical processes; one can opt for  
a preferred parametrization.

 A few words regarding the convergence of the expansion. We have already given the necessary condition in \eqref{eq:limit} when discussing the first order terms, but we wish to reassure the reader that the higher order terms respect it. In the above equations there are potentially large terms, that however by inspection are seen to be at most of order one
 \begin{align}
&\max\bigg| \frac{m_{d_{k_1}}}{m_{d_{k_1}}+m_{d_{j }}}  \bigg| \simeq 1 \\
&\max\bigg| \frac{m_{d_{k_1}}(m_{d_{k_1}}m_{d_{k_2}}-m_{d_{k_j}} m_{d_{k_3}})}{(m_{d_{k_1}}+m_{d_{j }})(m_{d_{k_1}}+m_{d_{k_3}})(m_{d_{k_2}}+m_{d_{j }})} \bigg|  \simeq 1
\end{align}
and thus the convergence is guaranteed when \eqref{eq:limit} is satisfied.
 The convergence of the series  was also checked by the comparison with the numerical results.

\section{Zero mixing angles}
\label{0mixing}
For the sake of illustration and simplicity, we go to the limit of vanishing LH mixing angles, which imply also diagonal $V_R$. This is useful for the case of small mixings, such as $\theta_{23}$ and  $\theta_{13}$ and gives an idea of what happens in the general case with non-zero angles. From \eqref{eq:VR}, one then finds  $V_R$ diagonal
with
\begin{align}\label{eq:exact}
(V_R)_{ii}= 
 \dfrac{ \pm \sqrt{1- \bigg(s_at_{2\beta}\dfrac{m_{u_i}^2-m_{d_i}^2}{2m_{d_i}m_{u_i}} \bigg)^2} -i s_at_{2\beta} \dfrac{ m_{u_i}^2+m_{d_i}^2}{2m_{d_i}m_{u_i}}   } {   1+i s_at_{2\beta}\,t_{\beta}e^{-ia} } 
\end{align}
The solution is valid when
\begin{equation}
 |s_at_{2\beta}|<  2\frac{|m_{d_{i}}m_{u_{i}}|}{|m_{u_{i}}^2-m_{d_{i}}^2| } 
\end{equation}
where one should  take clearly the smallest value.
 Applied,  for example, to the 2-3 generation case, this would give again roughly the limit $2 m_b/m_t$ found before.  
 In this case, by expanding in $s_at_{2\beta}$ one gets (ignoring corrections up to $m_b/m_t$)
 \begin{equation}
 (V_R)_{33}\simeq \pm\sqrt{1-\left( s_at_{2\beta} \frac{1}{2}\frac{m_{t}}{m_{b}}\right)^2     }-i s_at_{2\beta} \frac{1}{2}\frac{m_{t}}{m_{b}}
\end{equation}
   Since in our convention $(V_R)_{33}=e^{i\omega_3}$, this leads to
\begin{equation}  \label{eq:sinw3}
   \sin \omega_3\simeq -s_a t_{2\beta}  \frac{1}{2}\frac{m_{t}}{m_{b}}
\end{equation}  
As we show in  \eqref{eq:w3summed}, one gets basically the same result in the limit when the mixing angles are small.


\end{document}